\RequirePackage[2020-02-02]{latexrelease} 

\documentclass[aip,reprint,floatfix]{revtex4-1}

\usepackage{bm}
\usepackage{amsmath}
\usepackage{amssymb}
\usepackage{graphicx}

\draft 

\begin{document}

\title{K-space interpretation of image-scanning-microscopy}
\author{Tal I. Sommer}
\affiliation{Department of Applied Physics, Hebrew University of Jerusalem, Jerusalem 9190401, Israel}
\affiliation{Alexander Grass Center for Bioengineering, Hebrew University of Jerusalem, Jerusalem 9190401, Israel}

\author{Gil Weinberg}
\affiliation{Department of Applied Physics, Hebrew University of Jerusalem, Jerusalem 9190401, Israel}

\author{Ori Katz}
\affiliation{Department of Applied Physics, Hebrew University of Jerusalem, Jerusalem 9190401, Israel}

\date{\today}

\begin{abstract}
In recent years, image-scanning microscopy (ISM, also termed pixel-reassignment microscopy) has emerged as a technique that improves the resolution and signal-to-noise compared to confocal and widefield microscopy by employing a detector array at the image plane of a confocal laser scanning microscope. Here, we present a k-space analysis of coherent ISM, showing that ISM is equivalent to spotlight synthetic-aperture radar (SAR) and analogous to oblique-illumination microscopy. This insight indicates that ISM can be performed with a single detector placed in the k-space of the sample, which we numerically demonstrate.
\end{abstract}

\maketitle
 
Confocal imaging is a technique that improves axial-sectioning and transverse resolution in optical microscopy compared to conventional 'widefield' imaging\cite{mertz2019microscopy, sheppard2020reeval}. In widefield imaging, the entire field of view (FOV) is illuminated, and all points in the sample are simultaneously imaged to a detector (camera) plane. 
Confocal imaging allows an improvement in the transverse and axial resolution by illuminating the sample with a scanned focused spot, and detecting the signal emerging from that focal spot with a small 'confocal' detector conjugated to the illumination spot. 
The effective size of the confocal detector, which determines the spatial filtering and collection efficiency, is set by a small 'confocal pinhole' placed in front of the detector.
While the resolution improvement in confocal-microscopy increases with decreasing pinhole size\cite{fujimoto2009biomedical}, a smaller pinhole lowers the detection efficiency, resulting in a lower signal-to-noise ratio (SNR). 

A recently-introduced method termed image-scanning microscopy\cite{muller2010ism, ward2017image, tenne2019super} (ISM) or pixel-reassignment\cite{sheppard1988optik} allows full use of all signal photons in a confocal scanning system without sacrificing transverse imaging resolution. Moreover, depending on the imaging point spread function (PSF), ISM can also provide an improvement in imaging resolution\cite{sheppard2013superresolution, sheppard2013opra}. 
ISM achieves this feat by using an array of detectors at the image plane instead of the conventional single detector of confocal systems. The center detector of the ISM array collects the same information as would have been collected by a confocal pinhole. However, the neighboring detectors collect light that is otherwise rejected in a confocal system. To construct the ISM image, at each illumination point, $\bm{r}_{in}$, the signals from each detector position, $\bm{r}_{out}$, are reassigned to the midpoint between illumination and detection positions\cite{sheppard2020reeval, muller2010ism}: $\bm{r}_{ISM}=\frac{1}{2} (\bm{r}_{in}+\bm{r}_{out})$. The reassigned signals are summed over all scan positions, forming an ISM image. ISM was first implemented in incoherent microscopy modalities \cite{muller2010ism, winter2014two, roider2016high}, and was recently adapted to coherent imaging modalities \cite{bo2018pixel, dubose2019super,sommer2021upr, oron2022coherentISM}.

\begin{figure*}[htb!]
    \centering
    \includegraphics{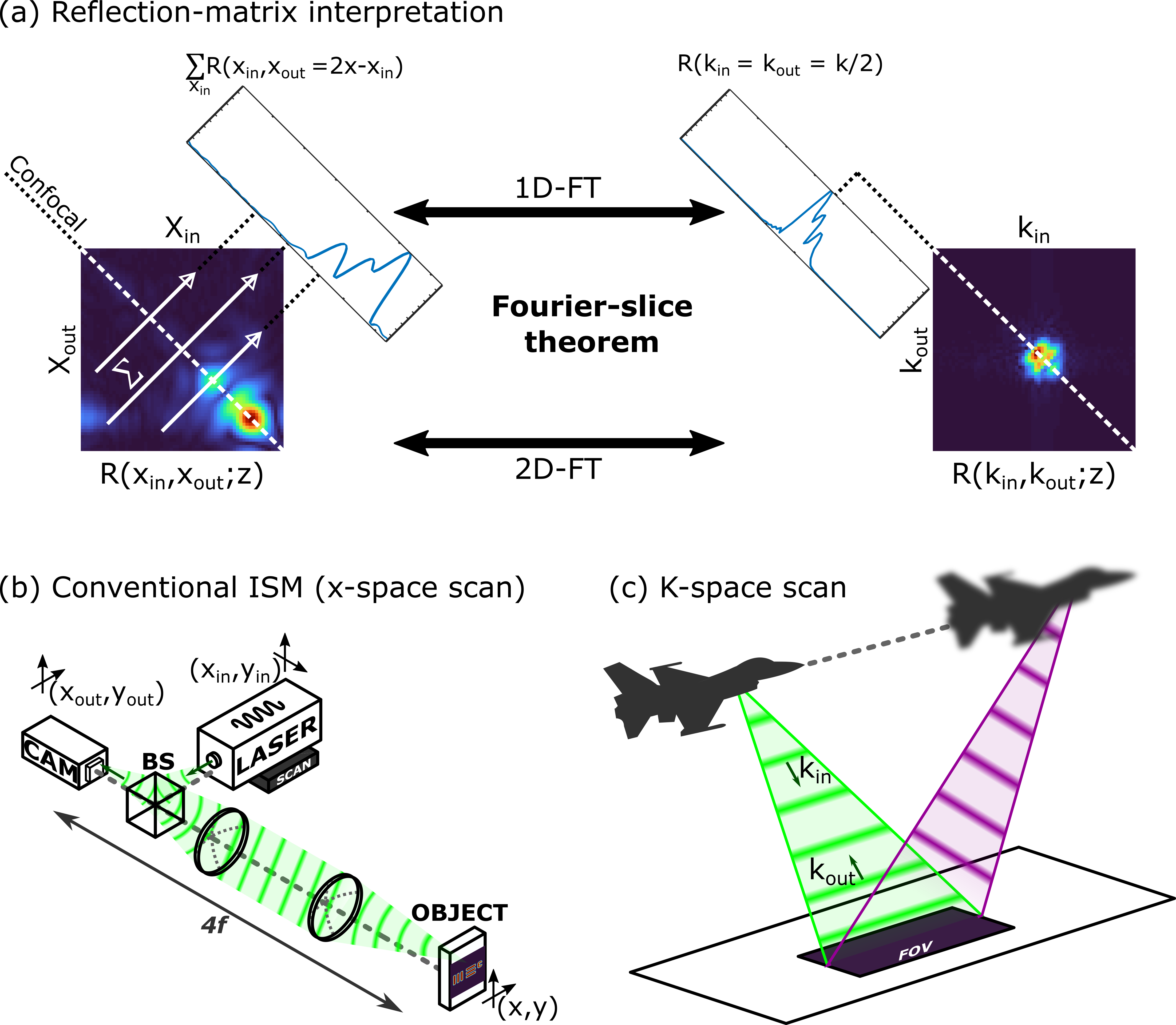}
    \caption{\label{fig:method} Coherent Image-Scanning-Microscopy (ISM), obtained by raster-scanning a focused illumination, and its k-space interpretation as spotlight synthetic aperture radar (SAR), obtained by plane wave illumination.
    (a) ISM measurements can be represented by a reflection matrix $R(x_{in},x_{out})$ providing the measured signals at coordinates $x_{out}$ for an illumination focused at $x_{in}$ (where the diagonal $R(x,x)$ is the confocal image). For simplicity, a one-dimensional object and monochromatic illumination are considered. 
    An ISM image, $I_{ISM}(x)$, is obtained by summing over the anti-diagonal elements of the reflection matrix, $I_{ISM}(x)=\sum_{x_{in}} R(x_{in},x_{out}=2x-x_{in})$, i.e., for all points satisfying $x=(x_{out}+x_{in})/2$  (white arrows). 
    According to the Fourier-slice theorem, this line summation (projection) is the Fourier-pair of the diagonal of the k-space reflection matrix $\tilde{R}(k_{in},k_{out})=\mathcal{F}\{R(x_{in},x_{out})\}$, where $\mathcal{F}\{\}$ is a 2D Fourier-transform. 
    These k-space equivalent measurements, $\tilde{R}(k_{in},k_{out}=k_{in})$, represent the reflected amplitude of a plane-wave at $k_{out}$ obtained for a plane-wave illumination at $k_{in}=k_{out}$, for a set of different illumination angles. These are analogous to spotlight SAR measurements.
    (b) Schematic of conventional ISM measurement setup: focused illumination is raster-scanned across the object plane (x-space). At each illumination position, the reflected fields at all positions around the illumination point are measured. 
    (c) Schematic of a setup for k-space ISM (analogous to spotlight SAR measurements): At each measurement position (green and purple beams), the object is illuminated by a tilted plane wave, $\bm{k}_{in}$, and the plane-wave that is reflected in the same angle, $\bm{k}_{out}=\bm{k}_{in}$, is detected. 
    }
\end{figure*}

Here, we analyze coherent-ISM in the spatial Fourier domain. Utilizing the projection-slice theorem\cite{bracewell1956Fslice_orig} we show that ISM is equivalent to spotlight synthetic-aperture radar (SAR) \cite{brown1967sar, kirk1975spot_sar} (Fig.~\ref{fig:method}), a well-established beam-scanning imaging technique, which utilizes a single detector. As a direct result, we numerically demonstrate that ISM can be performed with a single detector, and leverage the k-space analysis to highlight the close connection to oblique-illumination microscopy \cite{chowdhury2012oblq}.

We begin our analysis by representing the signals collected in coherent-ISM using the reflection matrix formalism\cite{aubry2020refmat, sommer2021upr}. In this formalism, the fields that are collected at position $\bm{r}_{out}$ at the detection plane when the illumination is focused at $\bm{r}_{in}$ in the object plane are given by the matrix element $R(\bm{r}_{in},\bm{r}_{out})$ (see coordinate notation in Fig.~\ref{fig:method}b). To simplify the mathematical derivations and without loss of generality, we consider below one transverse dimension, whose coordinate is given by $x$. 

Following the pixel reassignment process ($x_{ISM}=\frac{1}{2} (x_{in}+x_{out})$), the ISM image formation at a given imaging depth, $z$, is given by:

\begin{equation}
    I_{ISM}(x) = \displaystyle\sum_{x_{in}} R(x_{in},x_{out} = 2x - x_{in}; z)
\label{eq:ism_formation}
\end{equation}

In this matrical representation, the ISM summation can be interpreted as a sum over the anti-diagonal elements of the reflection matrix (Fig.~\ref{fig:method}a).

Such a line summation (Fig.~\ref{fig:method}a, left inset) is analogous to a line projection performed in many tomographic imaging techniques such as x-ray CT \cite{mersereau1974CT}. Leveraging the Fourier-slice theorem \cite{bracewell1956Fslice_orig, zhao1995Fslice_ct}, the Fourier transform of this line-projection is the main diagonal in the 2D Fourier transform of $R(x_{in},x_{out};z)$, scaled by a factor of half (Fig.~\ref{fig:method}a, right inset):

\begin{equation}
\mathcal{F}\{I_{ISM}(x)\}=\tilde{I}_{ISM}(k)=\tilde{R}(k_{in}=k/2,k_{out}=k/2;z)
\label{eq:ism_in_kspace}
\end{equation}

where $\mathcal{F}\{\}$ is the spatial Fourier-transform, and $k_{in}$ and $k_{out}$ are the k-vectors. For monochromatic illumination, this k-space reflection matrix\cite{aubry2020refmat_fourier}, $\tilde{R}(k_{in},k_{out})$, can be interpreted as the measured reflected plane-wave with a wavevector $k_{out}$ when illuminating the sample with a plane-wave, $k_{in}$. This k-space matrix is the 2D Fourier transform of $R$:

\begin{eqnarray}
    &&\tilde{R}(k_{in},k_{out};z) = \nonumber \\
    &&\iint{R(x_{in},x_{out}; z) ~e^{-ik_{in}x_{in}} ~e^{-ik_{out}x_{out}} ~dx_{out}dx_{in}} \nonumber \\
\label{eq:Rkk_def}
\end{eqnarray}

The Fourier-slice equivalence of Eq.~\ref{eq:ism_in_kspace} can be derived by considering the rescaled diagonal of $\tilde{R}$:

\begin{eqnarray}
    &&\tilde{R}(k_{in}=k/2,k_{out}=k/2;z) = \nonumber \\
    &&\iint{R(x_{in},x_{out}; z) ~e^{-i\frac{1}{2}kx_{in}} ~e^{-i\frac{1}{2}kx_{out}} ~dx_{out}dx_{in}} \nonumber \\
    &&=\iint{R(x_{in},x_{out}; z) ~e^{-ik\frac{x_{in}+x_{out}}{2}} ~dx_{out}dx_{in}}
\label{eq:Rkk_diag}
\end{eqnarray}

Performing inverse Fourier-transform yields:

\begin{eqnarray}
    &&\int{\tilde{R}(k_{in}=k/2,k_{out}=k/2;z) ~e^{ikx} ~dk} = \nonumber \\
    &&\iiint{R(x_{in},x_{out}; z) ~e^{-ik\frac{x_{in}+x_{out}}{2}} ~e^{ikx} ~dkdx_{out}dx_{in}} \nonumber \\
    =&&\iint{R(x_{in},x_{out}; z) ~\delta \left( x - \frac{x_{in}+x_{out}}{2} \right) ~dx_{out}dx_{in}} \nonumber \\
\label{eq:Rkk_diag_f}
\end{eqnarray}

Which is the same anti-diagonal summation as in ISM image formation in Eq.~\ref{eq:ism_formation}.


\begin{figure*}[htb!]
    \centering
    \includegraphics{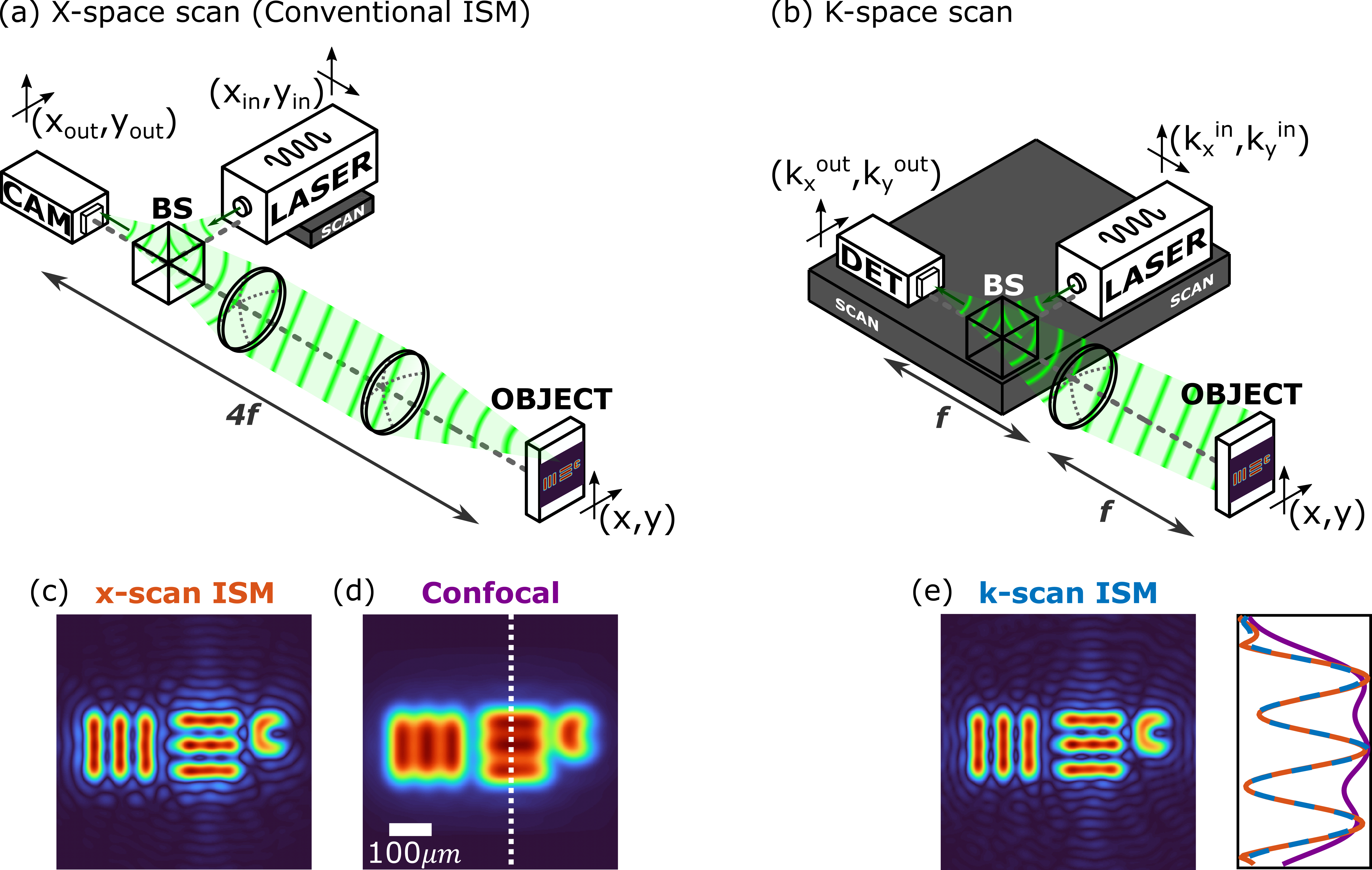}
    \caption{\label{fig:sim} Numerical comparison of coherent monochromatic imaging via conventional ISM, plane-wave scanning ISM, and confocal imaging.
    (a) In conventional ISM, a focused illumination beam raster scans the target object plane $(x,y)$. A detector array (CAM) that is conjugated to the object plane detects the reflected fields, and the ISM image is formed by reassignment and summation of the detected signals. A confocal image is formed from the signal emerging only from the illumination point ($(x_{in}, y_{in}) = (x_{out}, y_{out})$).
    (b) Monochromatic ISM can be performed with a single detector placed at the Fourier plane of the object, by sequentially illuminating the object with tilted plane waves.
    (c-e) We display a comparison between the detected amplitudes in confocal detection, an x-scanned ISM, and a k-scanned ISM. A vertical cross-section (depicted by a white dashed line in the confocal image) is also displayed. This comparison shows the equivalence between x-scan ISM and k-scan ISM results, which both result in a super-resolved image compared to the conventional confocal image.
    }
\end{figure*}

This mathematical equivalence implies that the same ISM image can be acquired in two different forms. One, conventionally, utilizes a scanning array in the spatial domain, detecting the reflected fields from a spatially-focused scanning illumination beam (Fig.~\ref{fig:method}b). Alternatively, the sample can be scanned by plane waves at different angles of incidence (given by $k_{in}$), and a single detector placed in the far field of the sample that detects the reflected plane wave at the same angle, i.e., the single Fourier component $k_{out}=k_{in}$.
This second approach, illustrated in Fig.~\ref{fig:method}c, is known as spotlight synthetic aperture radar (SAR) imaging \cite{kirk1975spot_sar, soumekh1999sar_processing}.

The scanning plane wave illumination of k-space ISM is closely related to oblique-illumination microscopy \cite{chowdhury2012oblq}. In oblique-illumination microscopy (originally proposed by Abbe\cite{abbe1873oblq}) the sample is illuminated by different angled plane-waves, and a camera is measuring the resulting image. In the k-space, the plane-wave illumination is manifested as a shift of the high frequencies angular-spectrum information of the sample into the passband of the system, allowing its detection \cite{wicker2014resolving}.
In k-space ISM, each illumination is identical to the plane wave illumination of oblique-illumination microscopy. The difference from oblique illumination microscopy is that only a single Fourier component is detected in each illumination in k-space ISM (Eq.~\ref{eq:ism_in_kspace}). Similar to oblique illumination microscopy, k-space ISM measurements thus yield extended k-space support. The differences between the techniques are the use of a single detector in k-space ISM rather than the detector-array of oblique-illumination microscopy, which requires a larger number of illuminations, and the angular-spectrum reweighting that is required in oblique illumination microscopy \cite{chowdhury2012oblq, ilovitsh2018structured}.

To support and demonstrate our mathematical derivation, we numerically simulated the coherent imaging of a test sample with conventional ISM (Fig.~\ref{fig:sim}a), and k-space ISM (Fig.~\ref{fig:sim}b). While the two approaches differ in both their illumination scheme (focused vs. plane-wave), and detection scheme (array detection in the object plane vs. a single detector in the k-space), the resulting images are identical and present an improvement over confocal imaging.

The simulated conventional ISM setup (Fig.~\ref{fig:sim}a) consists of a 4-f imaging system, where the scanning laser and the detection array (CAM) are both imaged onto the object plane.
A focused illumination beam raster scans the target object plane $(x,y)$. At each illumination position, $(x_{in},y_{in})$, the reflected field across the image plane, $(x_{out},y_{out})$, are measured. The simulated k-space ISM setup (Fig.~\ref{fig:sim}b) is based on Fourier-transforming a point illumination (LASER) and confocal detection, from the detector plane (DET) to the object plane using a lens. The point illumination is thus converted to a tilted plane-wave in the object plane, 
and the detector measures a single Fourier component, $k_{out}=k_{in}$. Observing Fig.~\ref{fig:sim}b reveals that k-space ISM is, in fact, confocal coherent imaging performed in the k-space domain. See simulation parameters in Supplementary Material, Section A.

In Fig.~\ref{fig:ultrasound} we compare the two ISM processing approaches and confocal imaging using an experimental ultrasound echography dataset. 
The dataset represents measurements performed on an acoustic phantom with multi-plane-wave transmission \cite{fink2009coherentcomp}. Data were acquired using a Verasonics P4–2v probe at a center frequency of $f_0=4.8MHz$, having $64$ elements with a total aperture size of $D=19.2mm$. The imaged target is composed of five pins (Fig.~\ref{fig:ultrasound}a, red X's) having a diameter of $0.1mm$ at a depth of $60mm$. 
Data were post-processed with the proper phases to perform coherent compounding for either confocal, x-space ISM\cite{sommer2021upr} or k-space ISM (Fig.~\ref{fig:ultrasound}a,b,c, respectively). These experimental results agree with the analytic and numerical investigations (Fig.~\ref{fig:ultrasound}d). See a discussion of the data processing in Supplementary Material, Section B.

\begin{figure}[htb!]
    \centering
    \includegraphics{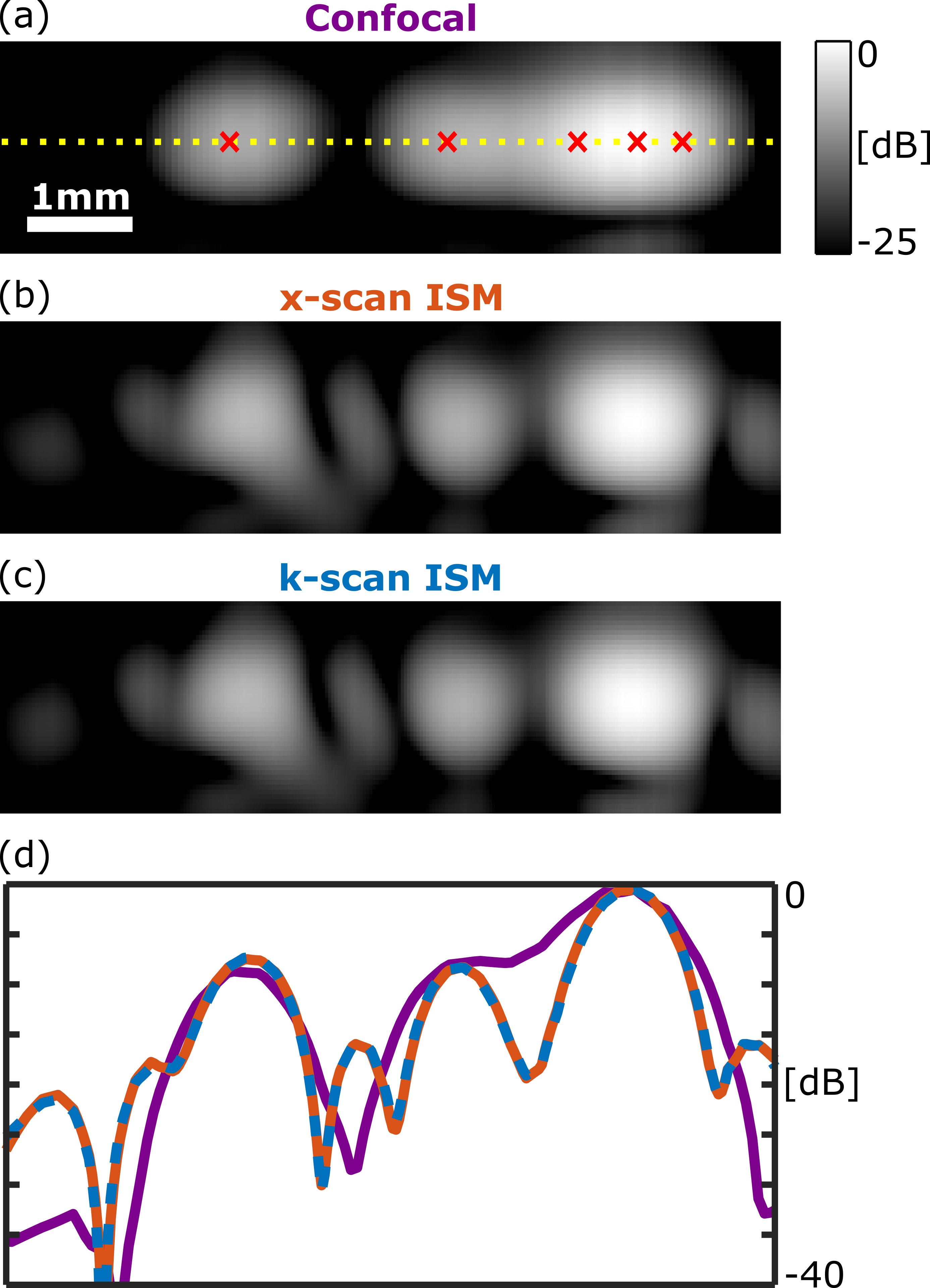}
    \caption{\label{fig:ultrasound} Experimental comparison of confocal ultrasound imaging (a),  conventional ISM (b), and k-space ISM (c), of five reflecting pins (marked by red X's) in an acoustic phantom. The same experimental data, acquired by plane wave sonications, were processed offline to generate the results. (a-c) presents the reconstructed intensity (absolute value square of the reconstructed fields).
    The two approaches for ISM results in practically identical images, having the expected noticeable resolution improvement over the confocal image.
    (d) A cross-section comparison of (a-c) along the dotted line in (a).
    }
\end{figure}


In conclusion, by leveraging the Fourier slice theorem, we show that it is possible to perform coherent ISM with a single detector located at the Fourier plane of the object. This allows the same resolution improvement gained in ISM over confocal imaging without requiring a detector array. This may be interesting for utilizing ISM in new modalities where detector arrays are not accessible.
In addition, the k-space acquisition and simple (confocal-like) analysis reduce the memory requirements and may be important also for the reduction of computational burden. 

We note that a single detector approach will result in a lower SNR than the detector-array approach in most scenarios since only a fraction of the reflected field is captured.
On the other hand, the plane-wave illumination reduces the power density concentrated on the sample, as compared to focused scanned illumination, and thus may allow higher illumination power.
 
\begin{acknowledgments}
This work has received funding from The Israel Science Foundation (Grant no. 1361/18), European Research Council (ERC) Horizon 2020 research and innovation program (677909), and was supported by the Ministry of Science and Technology in Israel.
\end{acknowledgments}

\section*{Author Contributions}
Tal Sommer and Gil Weinberg contributed equally to this work.

\textbf{Tal Sommer:}
Conceptualization (equal);
Data Curation (equal);
Formal Analysis(equal);
Methodology (equal);
Writing/Original Draft Preparation (lead);
Writing/Review \& Editing (equal).
\textbf{Gil Weinberg:}
Conceptualization (equal);
Data Curation (equal);
Formal Analysis(equal);
Methodology (equal);
Writing/Review \& Editing (equal).
\textbf{Ori Katz:}
Supervision (lead);
Funding Acquisition (lead);
Writing/Review \& Editing (equal)

\section*{Data availability}
The data that support the findings of this study are available from the corresponding author upon reasonable request.

\section*{Supplementary Material}
See supplementary material for the simulation parameters and a discussion of the data processing.

\bibliography{kISM}

\clearpage{}

\section*{Supplementary Material}

\subsection{\label{app:SIM_PARAMS}Simulation Parameters}
This section discusses the simulation parameters used to simulate the x-scanning and k-scanning systems (Fig.~2a-b in the main text).

Both systems were simulated using angular spectrum as a field propagator \cite{goodman2005fourier_optics}, and the lenses were simulated as a thin parabolic phase mask.
The illumination was set to a wavelength of $450nm$, and the object was imaged onto the detection plane using lenses of $35mm$ focal length and $~3mm$ in diameter. The transfer-function extent in the x-scanning system was controlled via an aperture in the Fourier-plane of the object. The aperture size is chosen such that the imaging system was of a numerical aperture (NA) of $0.0045$. This extent was preserved in the k-scanning system using a finite-extent scan of the Fourier-plane of the object to the same effective NA. 

\subsection{\label{app:US_BF}Ultrasound beam-forming}
This section discusses the data processing used for image reconstruction from an experimental ultrasound echography dataset (Fig.~3 in the main text).

The dataset represents measurements performed on an acoustic phantom (GAMMEX SONO403) with multi-plane-wave transmissions. Data were acquired using a Verasonics P4–2v probe at a center frequency of $f_0=4.8MHz$, having $64$ elements with a total aperture size of $D=19.2mm$, connected to a Verasonics Vantage 256 multi-channel system. The angular range of the steered plane-waves was determined by the probe's geometry via\cite{fink2009coherentcomp}:

\begin{equation}
    \theta_m = \arcsin{\frac{m\lambda_0}{D}}
\end{equation}

Where $\lambda_0$ is the probe's center wavelength.

These measurements can be represented with a matrix: $RF( u, t, \theta )$ where $u$ is the spatial position of the transducer element on the probe, $t$ is the measurement time, and $\theta$ is the steering angle of the transmitted plane-wave \cite{fink2009coherentcomp}.
Therefore, $RF( u, t, \theta )$ is the field measured at a transducer in position $u$ at time $t$ when sonicating with a plane-wave tilted with an angle $\theta$.
The temporal Fourier-transform of this matrix is: $RF(u,\omega,\theta)$.

Beamforming is done using the time-of-flight (TOF) for transmission (tx) and detection (rx) for each point in the imaged medium ($x$). The transmission time depends on the steering angle of transmission, and the detection time depends on the relative position of the transducer from the point. Assuming a single dimension, for simplicity, this TOF is: $t(x; u,\theta) = t_{tx}(x,\theta) + t_{rx}(x,u)$. Therefore, the reconstructed confocal image at position $x$ is a summation over all measurements at the respective TOF.

\begin{eqnarray}
    \label{seq:conf_img}
   &&I_{conf}(x) = \iint{RF (  u,t_{tx}(x,\theta)+t_{rx}(x,u), \theta ) du d\theta} \nonumber \\
    &&= \iint{RF (  u,t, \theta ) ~ \delta \left(t-\left(t_{tx}(x,\theta)+t_{rx}(x,u)\right) \right) du d\theta} \nonumber \\
\end{eqnarray}

A reflection matrix (that is used for x-scan ISM image formation) can be constructed by using a transmission time for a position that is different from the position for the detection time \cite{aubry2020refmat,aubry2020refmat_fourier}:

\begin{eqnarray}
    \label{seq:Rmat}
    &&R(x_{in},x_{out}) = \iint{RF \left(  u,t_{tx}(x_{in},\theta)+t_{rx}(x_{out},u), \theta \right) du d\theta} \nonumber \\
    &&= \iint{RF (  u,t, \theta ) ~ \delta \left(t-\left(t_{tx}(x_{in},\theta)+t_{rx}(x_{out},u)\right) \right) du d\theta} \nonumber \\
\end{eqnarray}

This operation is termed Delay-And-Sum (DAS).

In the temporal Fourier-domain, these beamformations can be performed using phases instead of time-delays:

\begin{subequations}

    \begin{equation} \label{seq:conf_img_w}
        I_{conf}(x) = \iiint { RF(u,\omega,\theta) e^{i \omega \left(t_{tx}(x,\theta)+t_{rx}(x,u)\right)}} du d\theta d\omega
    \end{equation}

    \begin{equation}\label{eq:Rmat_w}
        R(x_{out}, x_{in}) = \iiint { RF(u,\omega,\theta) e^{i \omega \left(t_{tx}(x_{in},\theta)+t_{rx}(x_{out},u)\right)}} du d\theta d\omega
    \end{equation}
    
\end{subequations} 

The k-space reflection-matrix, $\tilde{R}(k_{in},k_{out})$, is the 2D spatial Fourier transform of $R(x_{in},x_{out})$:

\begin{equation}
\label{seq:Rkk_def}
    R(k_{out}, k_{in}) = \iint { R(x_{out}, x_{in}) e^{-i k_{in} x_{in}} e^{-i k_{out} x_{out}} dx_{in} dx_{out}}
\end{equation}

And, as the $RF$ measurement is not dependant of $x_{in}$ and $x_{out}$, Eq.~\ref{seq:Rkk_def} can be written as:

\begin{eqnarray}
\label{seq:Rkk_1}
    R(k_{out}, k_{in}) &&= \iiint RF(u,\omega,\theta) \nonumber \\
    \times &&\left(\int{e^{i \omega t_{tx}(x_{in},\theta)}e^{-i k_{in} x_{in}}dx_{in}}\right) \nonumber \\
    \times &&\left(\int{e^{i \omega t_{rx}(x_{out},\theta)}e^{-i k_{out} x_{out}}dx_{out}}\right)du d\theta d\omega \nonumber \\
\end{eqnarray}

This means that the k-space reflection-matrix can be constructed in the temporal Fourier-domain, using the 1D spatial Fourier-transforms of the phases used for the reflection-matrix construction.

One should determine the field-of-view (FOV) lateral grid to be the same as the positions of the transducers ($u$). This results in the grids of the two 1D spatial Fourier-transforms to coincide. One can leverage this method to reconstruct a k-scanned ISM image directly from the echography dataset using only the phases that correspond to the wave-vectors $k_{in}=k_{out}$.


\end{document}